\documentstyle{mn}
\input epsf.sty
\newif\ifAMStwofonts
\AMStwofontstrue
 
\def\rmax{R_{\rm max}}
\def\RSch{R_{\rm Schw}}
\def\MEdd{\dot M_{\rm Edd}}
\def\Pgas{P_{\rm gas}}
\def\Fc{F_{\rm c}}
\def\Sigmac{\Sigma_{\rm c}}
\def\Fadv{F_{\rm adv}}
\def\Fsoft{F_{\rm soft}}
\def\Mc{\dot M_{\rm c}}
\def\Md{\dot M_{\rm d}}
\def\Te{T_{\rm e}}
\def\Ts{T_{\rm s}}
\def\Ti{T_{\rm i}}
\def\Tv{T_{\rm vir}}
\def\taues{\tau_{\rm es}}
\def\mdadv{\dot m_{\rm adv}}
\def\MSun{{\rm M}_{\sun}}
 
\ifoldfss
  \ifCUPmtlplainloaded \else
    \NewTextAlphabet{textbfit} {cmbxti10} {}
    \NewTextAlphabet{textbfss} {cmssbx10} {}
    \NewMathAlphabet{mathbfit} {cmbxti10} {} % for math mode
    \NewMathAlphabet{mathbfss} {cmssbx10} {} %  "   "    "
  \fi
  \ifAMStwofonts
    \ifCUPmtlplainloaded \else
      \NewSymbolFont{upmath} {eurm10}
      \NewSymbolFont{AMSa} {msam10}
      \NewMathSymbol{\upi}     {0}{upmath}{19}
      \NewMathSymbol{\umu}     {0}{upmath}{16}
      \NewMathSymbol{\upartial}{0}{upmath}{40}
      \NewMathSymbol{\leqslant}{3}{AMSa}{36}
      \NewMathSymbol{\geqslant}{3}{AMSa}{3E}

       \let\le=\leqslant
       \let\ge=\geqslant
    \fi
  \fi
\fi % End of OFSS
 
\ifnfssone
  \newmathalphabet{\mathit}
  \addtoversion{normal}{\mahit}{cmr}{m}{it}
  \addtoversion{bold}{\mathit}{cmr}{bx}{it}
  \newmathalphabet{\mathbfit} % math mode version of \textbfit{..}
  \addtoversion{normal}{\mathbfit}{cmr}{bx}{it} 
  \addtoversion{bold}{\mathbfit}{cmr}{bx}{it}
  \newmathalphabet{\mathbfss} % math mode version of \textbfss{..}
  \addtoversion{normal}{\mathbfss}{cmss}{bx}{n}
  \addtoversion{bold}{\mathbfss}{cmss}{bx}{n}
  \ifAMStwofonts
    \ifCUPmtlplainloaded \else
and
your
      \UseAMStwoboldmath
      \makeatletter
      \new@mathgroup\upmath@group
      \define@mathgroup\mv@normal\upmath@group{eur}{m}{n}
      \define@mathgroup\mv@bold\upmath@group{eur}{b}{n}
      \edef\UPM{\hexnumber\upmath@group}
      \new@mathgroup\amsa@group
      \define@mathgroup\mv@normal\amsa@group{msa}{m}{n}
      \define@mathgroup\mv@bold\amsa@group{msa}{m}{n}
      \edef\AMSa{\hexnumber\amsa@group}  
      \makeatother
      \mathchardef\upi="0\UPM19
      \mathchardef\umu="0\UPM16
      \mathchardef\upartial="0\UPM40
      \mathchardef\leqslant="3\AMSa36
      \mathchardef\geqslant="3\AMSa3E

       \let\le=\leqslant
       \let\ge=\geqslant 
    \fi
  \fi
\fi % End of NFSS release 1
   
\ifnfsstwo
  \DeclareMathAlphabet{\mathbfit}{OT1}{cmr}{bx}{it}
  \SetMathAlphabet\mathbfit{bold}{OT1}{cmr}{bx}{it}
  \DeclareMathAlphabet{\mathbfss}{OT1}{cmss}{bx}{n}
  \SetMathAlphabet\mathbfss{bold}{OT1}{cmss}{bx}{n}
  \ifAMStwofonts
    \ifCUPmtlplainloaded \else
      \DeclareSymbolFont{UPM}{U}{eur}{m}{n}
      \SetSymbolFont{UPM}{bold}{U}{eur}{b}{n}
      \DeclareSymbolFont{AMSa}{U}{msa}{m}{n}
      \DeclareMathSymbol{\upi}{0}{UPM}{"19}
      \DeclareMathSymbol{\umu}{0}{UPM}{"16}
      \DeclareMathSymbol{\upartial}{0}{UPM}{"40}
      \DeclareMathSymbol{\leqslant}{3}{AMSa}{"36}
      \DeclareMathSymbol{\geqslant}{3}{AMSa}{"3E}

       \let\le=\leqslant
       \let\ge=\geqslant
    \fi
  \fi  
\fi % End of NFSS release 2
      
\ifCUPmtlplainloaded \else
  \ifAMStwofonts \else % If no AMS fonts
    \def\upi{\pi}
    \def\umu{\mu}
    \def\upartial{\partial}
  \fi
\fi

\title{The role of advection in the accreting corona model for active
galactic nuclei and Galactic black holes}
\author[A. Janiuk, P.T. \.{Z}ycki, B. Czerny]
       {A. Janiuk$^1$, P.T. \.{Z}ycki$^{2,1}$, B. Czerny$^1$\\
        $^1$N. Copernicus Astronomical Center, Bartycka 18, 00-716 Warsaw, 
        Poland\\$^2$ Department of Physics, University of Durham, Durham DH1
        3LE}
\begin{document}

\maketitle

\begin{abstract}

We consider the role of advection in the two--temperature accreting corona
with an underlying optically thick disc.
The properties of coronal solutions depend significantly on the description of
advection. Local parameterization of advection by a constant coeficient
$\delta$ replacing the radial derivatives lead to complex topology of 
solutions, similar to some extent to other advection-dominated accretion flow
solutions.  One, radiatively cooled 
branch exists for low accertion rates. For higher accretion rates two
solutions exist in a broad range of radii: one is radiatively cooled and the
other one is advection-dominated. With further increase of accretion rate the
radial extensions of the two solutions shrink and no solutions are found above
certain critical value. However, these trends change if the local 
parameterization of advection is replaced by proper radial derivatives 
computed iteratively from the
model. Only one, radiatively cooled solution remains, and it exists even
for high accretion rates. The advection-dominated branch disappears during the
iteration process which means that a self-consistently described 
advection-dominated flow cannot exist in the presence of an underlying cold
disc.

\end{abstract}


\section{Introduction}

There are direct observational evidences that the gas surrounding central 
black 
holes in active galactic nuclei (AGN) and in galactic black holes (GBH) 
consists
of two phases: colder, optically thick phase and hotter optically thin phase
(for a review, see Mushotzky, Done \& Pounds 1993; Tanaka \& Lewin 1995;
Madejski 1999). 
The emission of radiation is clearly powered by accretion onto 
the black hole. However, it is still not clear which of the two phases 
is responsible for accretion.

In the classical disc/corona models (e.g. Liang \& Price 1977;
Bisnovatyi-Kogan \& Blinnikov 1977; Haardt \& Maraschi 1991) 
the accretion proceeds through the disc and
the coronal gas does not contribute significantly to the angular momentum 
transfer, presumably because of its
magnetic coupling to the disc.

In clumpy accretion flow models the accretion proceeds predominantly through 
the cold clumps of gas (Collin--Souffrin et al.\ 1996, Czerny \& Dumont 1998; 
Krolik 1998; see also Torricelli-Ciamponi \& Courvoisier 1998).

However, there are also models in which the hot gas carries most of the mass.
The division of the flow into the two phases is predominantly radial
in models based on advection dominated accretion flow (ADAF) solutions 
(e.g.\ Ichimaru 1977; Abramowicz et al.\ 1995; Narayan, Kato \& Honma 1997;
Esin, McClintock \& Narayan 1997), in the Compton cooled solutions 
for ion tori (Shapiro,
Lightman \& Eardley 1976; hereafter SLE), or in models considering both 
cooling mechanism (Bj\"ornsson et al.\ 1996). In those models the accretion 
proceeds through the cold accretion disc in the outer 
region and through the hot plasma region in the inner region.

In this paper we discuss the model of an accreting corona. Initial formulation
of the basic assumptions was given by \. Zycki, Collin-Souffrin \& 
Czerny (1995) and the final formulation of the model was outlined by 
Witt, Czerny \& \. Zycki (1997; hereafter WCZ). 
In this model the accretion proceeds both 
through a disc and a corona, in  proportions determined by the model and 
varying with the distance from the black hole. The model was applied
to Nova Muscae 1991 (Czerny, Witt \& \. Zycki 1996) and 
AGN (Czerny, Witt \& \. Zycki 1997).

In this paper we address the problem of advection in the corona. An optically 
thin accreting corona should show similar general behavior as a general
optically thin flow not accompanied by the disc, i.e.\ we might expect both
advection dominated solutions and radiatively cooled solutions. 
Advection
was included in  the model by WCZ and was found to be always negligible. Here
we explain the nature of this phenomenon.

\section{Model}

\subsection{Corona structure}

We assume that the corona itself accretes, i.e.\ that the energy in the corona
is due to the direct release of the gravitational energy in the hot phase, 
without the necessity of having a mediator (e.g.\ magnetic field) between the 
disc and the corona. We further assume that the energy release can be
described by the $\alpha$ viscosity prescription of Shakura \& Sunyaev (1973),
with the energy generation rate proportional to the gas pressure (we neglect
radiation pressure in the corona because the corona is optically thin).

We assume a two-temperature plasma in the corona as in the classical paper 
of SLE,
i.e.\ the ion temperature is different from the electron
temperature. The energy balance is computed  assuming that the release of 
gravitational energy heats the ions, the Coulomb
coupling transfers this energy to electrons and finally electrons cool down
by inverse Compton process, with the disc emission acting as a source of soft
radiation flux.
The hot corona is  radiatively coupled to the disc, as described by 
Haardt \& Maraschi (1991). We assume isotropic emission within the corona
($\eta=1/2$)  and for the (energy integrated) disc albedo we adopt 
a value $a= 0.2$.
Compared to our previous paper (WCZ) we now use an accurate prescription 
for the Compton amplification factor, $A$ (defined by $\Fc = A\,\Fsoft$),
namely we compute $A$ from Monte Carlo simulations of Comptonization. 
Our Monte Carlo code employs the method described by Pozdnyakov, Sobol \& 
Sunyaev (1983) and G\'{o}recki \& Wilczewski (1984). Assuming slab geometry
(Thomson optical depth $\taues$ and electron temperature $\Te$)
and the soft photons spectrum as a black body of temperature $\Ts$, we compute
$A$ on a grid of $\Te,\ \taues,\ {\rm and}\,\Ts$ and interpolate for values
of interest at each radius.

We neglect the dynamical term in the corona since it was shown to be
relatively unimportant (WCZ). Therefore we can assume that the
corona is in hydrostatic equilibrium and instead of solving original complex 
differential equations we adopt a set of simplified equations given in
Appendix D of WCZ, with modifications concerning the advection and the Compton
amplification factor.
We also neglect the effect of the vertical outflow of the
gas from the corona discussed by WCZ, i.e.\ we assume that the total 
accretion rate 
(the sum of accretion rates through the disc and the corona) is constant,
independent of radius. 
%We include the effect of non-conservative accretion
%when stated specifically.

\subsection{Disc--corona transition}

The main feature of the model is that 
the division of the accretion flow into the hot and cold phases
is not arbitrary. 
The thermal instability in an irradiated medium (Krolik, McKee \& Tarter 1981)
results in its spontaneous stratification into two stable phases.
The criterium for  the transition from cold to hot phase is 
formulated in terms of a specific value of the ionization parameter $\Xi$
which we define as
\begin{equation}
\label{equ:xi}
\Xi \equiv {\eta\Fc \over c \Pgas}
\end{equation}
where $\eta\Fc$ is the fraction of the coronal flux directed towards the disc
and $\Pgas$ is the coronal gas pressure (see also Krolik 1998). 
Following Begelman, McKee \& Shields (1983) we adopt the following scaling,
\begin{equation}
\label{equ:xi2}
 \Xi = 0.65 (\Te/10^8\ {\rm K})^{-3/2}
\end{equation}
where $\Te$ is the electron temperature of the corona at a given radius.

This description of the disc/corona transition can be easily argued for in a 
qualitative way. The essence of the transition is a change from Compton 
cooling mechanism to atomic cooling, including bremsstrahlung. Therefore, 
the transition zone corresponds to certain assumed contribution of 
bremsstrahlung to the total cooling. The criterion (Eq.~\ref{equ:xi2}) applies 
accurately to the systems with either Compton heating of the corona or a 
heating proportional to the density and fixes the bremsstrahlung contribution 
to 2/3 of the total cooling. Similar criterion formulated by Krolik (1998) 
gives the value of 3/7. Since the relative bremsstrahlung contribution 
decreases rapidly with the departure from the transition zone into the corona,
while the pressure remains roughly constant, we use this criterion as 
the basic criterion for pressure and we 
neglect bremsstrahlung as a cooling mechanism (see also Krolik 1998). 

This additional equation enables us to actually determine the fraction 
of energy, $f$, which is liberated in the corona (N.B.\ in our previous 
paper WCZ we used $\xi\equiv 1-f$). Since in our model $f$ describes also
the fraction of the mass accreting
through the corona, we have
\begin{equation}
\Mc(r) = f(r) \dot M; ~~~~~~~~~~~~~\Md(r) = [1-f(r)] \dot M 
\end{equation}
where $\Md$, $\Mc$ and $\dot M$ are the disc, coronal and the total mass 
accretion rates. The total accretion rate does
not depend on $r$ if mass loss (e.g.\ through a wind as in WCZ) is neglected.

The formulation of the model allows for computing the corona structure 
independently form the disc internal
structure. The disc/corona coupling is through surface pressure, $\Pgas$,
the coronal radiation flux, $\Fc$ and the disc soft flux, $\Fsoft$ 
uniquely specified by $f$ and albedo. There is no need for subsequent 
iterations between the computations of the corona structure and the disc 
vertical structure as long as the local disc emission is well approximated 
by a blackbody (more generally: a thermal emission with a constant 
ratio of 
the colour and effective temperatures) and the albedo is fixed. 
When the local corona parameters  are determined, they 
provide the surface boundary conditions for the equations of the cold disc 
structure. The disc vertical structure can be solved if the viscous transfer
within the disc is specified (R\' o\. za\' nska et al. 1999). The imposed
boundary conditions are automatically satisfied by the model, and the disc
structure is determined uniquely.

 The radial variation of the relative proportion of the disc and coronal
accretion flows  may mean that either a fraction of the coronal mass
cools and settles down on the disc surface thus joining the disc flow
instead of falling radially into black hole, or a fraction of the disc flow 
evaporates from the disc surface and
joins the coronal flow. These changes are forced by the requirement
of the hydrostatic and thermal equilibrium between the disc and the corona at
each radius. However, the dynamics  of this phenomenon is beyond the scope
of our present model.

\subsection{Advection}

In the present paper we include the advection term in the corona in all
 computations. The advection is described as in WCZ, i.e.\ the energy
balance takes the form 
\begin{equation}
 \Fc=F_{\rm CC} + \Fadv
\end{equation}
and
\begin{equation}
\label{equ:delta}
4 \pi r^2 F_{adv}=f(r) \dot M c_s^2 \delta; \ \ \ \ 
\delta= {d \ln P \over d \ln r} - {5 \over 2} {d \ln \Ti \over d \ln r},
\end{equation} 
where $\Fc$ is the energy flux generated in the corona, $F_{\rm CC}$ is the
Compton cooling of the corona by the soft disc photons and $c_s$ is the
sound velocity in the corona. 
Equation~\ref{equ:delta} can be converted to give
\begin{equation}
\label{equ:advfrac}
{F_{\rm adv} \over \Fc} = \delta\, {\Ti \over \Tv},
\end{equation}
where $\Tv$ is the virial temperature, 
\begin{equation}
\Tv \equiv { G M \over r } {m_{\rm H} \over k}.
\end{equation}
For the numerical solution of WCZ, $\delta=0.75$ (see Appendix C in that 
paper). In general, however, $\delta$ is a function of radius and 
should be computed consistently. In the next Section we will show  solutions
for a number of fixed values of $\delta$, while in Section~\ref{sec:diter}
we will discuss solutions
with $\delta$ determined consistently through iterations.

The parameters of the model are: the viscosity parameter $\alpha$ in the 
corona and the dimensionless accretion rate $\dot m$ 
measured in the Eddington units 
\begin{equation}
\MEdd =3.52\,  { M \over 10^8 \MSun}\, [\MSun/yr]
\end{equation}
where $M$ is the mass of the central black hole and we assumed the
pseudo--Newtonian efficiency of accretion equal 1/16 and pure hydrogen
opacities. 

We show the results for the
value of the viscosity parameter $\alpha=0.3$ and the mass of the black hole
$M = 10 M_{\odot}$ but we discuss the 
trends of solutions with these parameters  varied. 

\subsection{Spectra}

At each radius, the equations of the structure of the corona determine the 
electron 
temperature, the optical depth of the corona and the soft photon flux 
from the disc. The effect of the Comptonization of the disc flux by the 
corona is calculated at each radius separately, using semi-analytical
formulae of Czerny \& Zbyszewska (1991).
We neglect here the anisotropy of the Compton scattering which should be 
taken into account in detailed spectral modeling (e.g.\ Poutanen \&
Svensson 1996, 
Haardt, Maraschi \& Ghisellini 1997). However, very accurate computation of 
the spectra  is not the main goal of the present paper.

The final disc spectrum is computed by integration over the disc surface
assuming an inclination angle equal zero (i.e. top view). This integration 
procedure is an essential element of our model since the corona properties are
radius--dependent: outer parts of the corona are predominantly responsible
for the high energy extension of the spectrum and its hard X-ray slope 
while the inner, cooler parts of the corona  mostly influence the soft X-ray 
range by producing moderately comptonized disc component.
 
All computations are
done for a non--rotating black hole and the relativistic corrections are
neglected.

\section{Corona properties for constant $\bmath{\delta}$}
\label{sect:dconst}

In this Section we show results obtained assuming a value of the advection 
parameter $\delta$ (Eq.~\ref{equ:delta}) constant with radius. 

\subsection{The relative strength of the corona}

\begin{figure*}
\epsfxsize = \textwidth
\epsfbox[40 150 560 700]{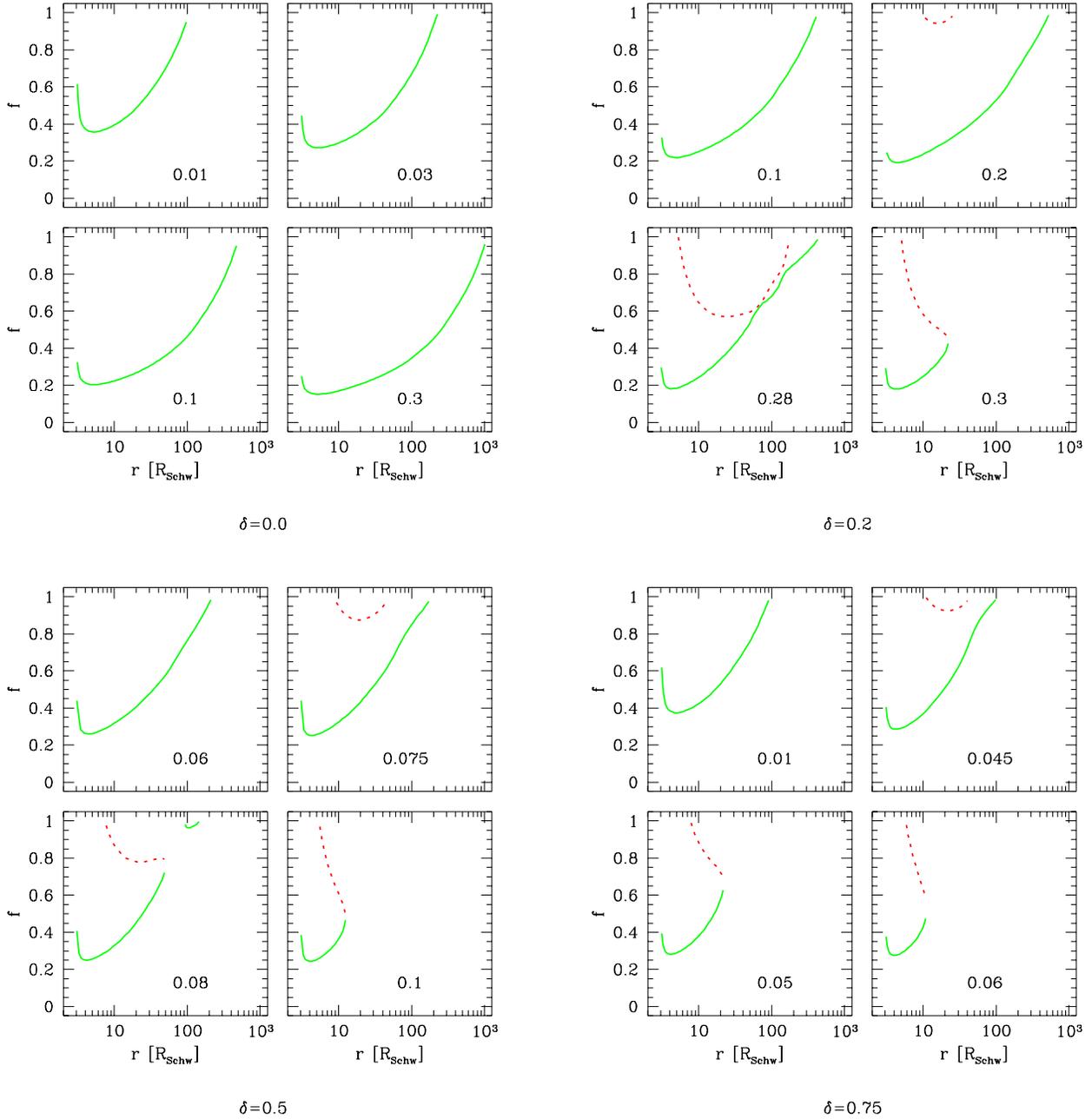}
\caption{The fraction of the energy dissipated in the corona, $f$, as a 
function of
radius for viscosity parameter $\alpha=0.3$, mass of the black hole 
$10 M_{\odot}$, and four values of the advection parameter $\delta$
(Equation~\ref{equ:delta}): $\delta=0,\ 0.2,\ 0.5\ {\rm and}\ 0.75$.
Labels inside panels refer to accretion rate.
\label{fig:xi}}
\end{figure*}

The most important prediction of the model is a strong 
radial dependence of the strength of the corona.
This dependence changes qualitatively with the accretion rate, 
mainly due to the presence of advection.
Examples of radial dependences of the fraction of energy generated in the
corona, $f(r)$, are shown in Figure ~\ref{fig:xi} for a number of values
of $\delta$ and $\dot m$.

Spatial extent of the corona is always finite, independently of $\dot m$
and $\delta$. The corona covers only an inner part of the disc from a certain 
outer radius $\rmax$ inwards. 

At $\rmax$ all the energy is
liberated in the corona  for low accretion rates, if $\delta \ne 0$ 
(and for any $\dot m$ when $\delta=0$, i.e.\ if there is no advection).
Consequently, at $r=\rmax$ the accretion flow proceeds entirely through 
the corona.
At larger radii no corona solutions of our equations exist since
the Compton cooling provided by the disc
is too large, under the adopted assumptions about the corona structure. 
There is therefore a strong and 
discontinuous change of accretion flow structure at $\rmax$. For
$r>\rmax$ all the accretion proceeds through the disc whilst 
at $r=\rmax$ the accretion proceeds through the corona, with the cold 
disc heated only by X-ray illumination. We envision that the change occurs
through rapid heating and evaporation of the disc surface which takes place
in order to maintain the thermal balance condition, but the detailed
dynamics of this process is beyond the scope of our present model.
Closer in, the relative strength of the
corona decreases and, consequently, the relative fraction of the disc accretion
increases.

\begin{figure}
    \epsfxsize = 80 mm 
\epsfbox[40 260 560 700]{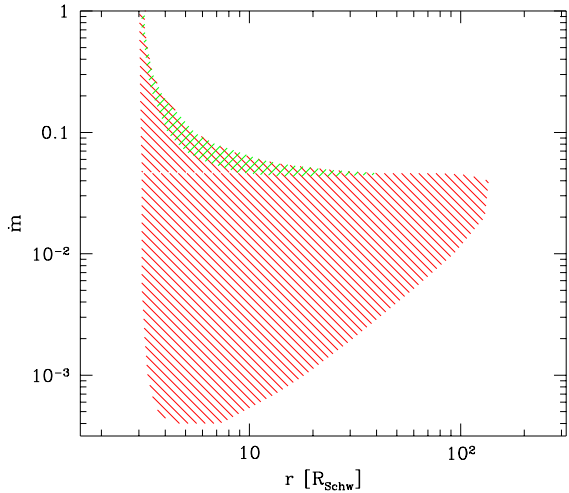}
\caption{The dependence of the extension of the corona on the accretion rate
for $\delta=0.75$.  Light shades mark the Compton
cooling dominated solutions and the dark shades mark the region of both
advection dominated and cooling dominated  solutions.
The viscosity parameter $\alpha=0.3$  and  the black hole  mass is
$10 \MSun$.
\label{fig:topo}}
\end{figure}

The dependence of the radial extension of the corona, $\rmax$, 
for $\delta=0.75$ on the accretion
rate is shown in Figure ~\ref{fig:topo}. The size of the corona is considerable, 
covering the disc up to $\sim 130 \RSch$ for accretion rate 
approaching $0.04\,\MEdd$, but $\rmax$
decreases significantly for smaller accretion rates, down to 
about $10 \RSch$ for sources radiating at 0.1 per cent of the Eddington 
luminosity. For accretion rates below $4 \times 10^{-4}\,\MEdd$ 
the corona ceases to exist, as the Compton cooling is too strong while heating
 becomes inefficient.
(see also Section ~\ref{sec:form}).

Although the relative fraction of energy dissipated in the corona 
increases 
with radius (up to $\rmax$) the actual energy flux decreases outwards. We can
see that from the simple analytical solutions given in Appendix C of WCZ. 
Since $f$  increases with the radius $r$ approximately as $f \propto r^{3/8}$,
the
X-ray flux $F_{\rm X}(r)$ decreases with radius: 
$F_{\rm X}(r) \propto  r^{-3+3/8}$.
This may be important for detailed computations of relativistic smearing
of the X-ray reprocessed component. However, we do not
discuss this spectral component in the present paper.

A second branch of solutions appears when the accretion rate $\dot m$ 
approaches a certain critical value, $\mdadv$. On this branch the cooling 
is dominated by advection and $\mdadv$ is a function of 
$\delta$: $\mdadv=0.192,\ 0.069,\ 0.044$ for
$\delta=0.2,\ 0.5,\ 0.75$, respectively (Figure ~\ref{fig:xi}).
The exact topology of the two solutions is a rather complex function
of $\delta$ and $\dot m$. For $\dot m$ just above $\mdadv$ the advective
solution coexists with the radiative one in a range of radii. The two
branches cross for somewhat higher $\dot m$ and then separate, creating 
two spatial regions where both solutions can exist, 
with an intermediate range of $r$, where
no solution is possible. The outer ring then shrinks rapidly as $\dot m$
increases and disappears, the inner one does the same but more slowly.
Mathematically, the solutions continue to (unphysical) $f>1$ region 
creating closed loops.

For comparison we also plot in Figure ~\ref{fig:xi} the radial dependence
of the relative strength of corona obtained when the advection term was 
neglected ($\delta=0$). 
Obviously, in this case only one branch of solutions appears.
It is similar to the Compton--cooled branches of solutions with $\delta>0$ 
for low accretion rates, $\dot m \la 0.01$, i.e.\ the advection is then only
a small correction to the energy balance. For
larger $\dot m$ the $\delta=0$ solution is quite different from 
the advection-corrected solutions: the corona extends
to much larger radii and no qualitative change with increasing $\dot m$
is seen. It is clearly the advection term  which is responsible for 
the complex topology of solutions seen in Figure ~\ref{fig:xi}.

\begin{figure}
\epsfxsize = 80 mm 
\epsfbox[30 300 600 680]{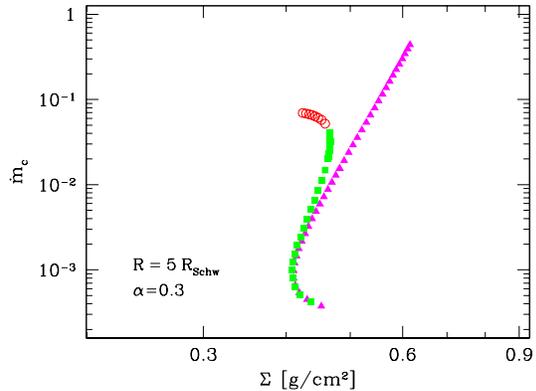}
\caption{Coronal solutions shown in coronal accretion 
rate, $\dot m_{\rm c}$, versus  surface density in the corona, $\Sigma$, 
computed at $5\, \RSch$ for $\alpha=0.3$.
Open circles show advection-dominated solution and solid squares show the 
Compton cooling dominated solution for $\delta=0.75$.
Note that the advective branch terminates when $F_{\rm adv}/\Fc=0.9$, 
i.e.\ strongly advective solutions do {\it not\/}  exist in our model. 
Solid triangles show the solution
with $\delta(r)$ computed consistently through iterations 
(Section~\ref{sec:diter}). Only one solution branch exists in this case,
and it extends into fairly high $\dot m_{\rm c}$ (corresponding to
total $\dot M \gg \MEdd$). Black hole mass $M = 10\, \MSun$. 
\label{fig:ADAF}}
\end{figure}

The merging of the radiatively cooled solution with the advectively cooled 
solution has been previously found by Chen et al.\ (1995) and 
Bj\" ornsson et al. (1996) for an optically thin flow, 
not accompanied by any disc.
%, studied at a given radius. 
Therefore, we also constructed an analogous $\log \dot m - \log \Sigma$
diagram for our coronal solution at $5\,\RSch$ (Figure ~\ref{fig:ADAF}).  
We plot coronal surface density $\Sigmac$ versus the coronal accretion rate, 
$\dot m_{\rm c} = \dot m\,f(r)$ where the value of factor $f(r)$ depends 
on the solution branch.

The two solution branches for a given total $\dot m$ produce two
points on this plot, since $f$ is different on the two branches.
The two solutions merge  for $\dot m = 0.09 $ 
(at our assumed radius $5\, \RSch$, $\alpha=0.3$ and $\delta=0.75$), 
resulting in a continuous curve with both the fraction of energy carried by
advection and the fraction of energy dissipated within the corona 
increasing with $f \dot m$. The uppermost point is characterized
by $f = 1$, i.e.\ the whole energy is generated within the corona, 
but advection transports $\approx 90$ per cent of the energy rather than
100 per cent. In our model the fully advective branch  does not develop,  
due to the presence of the disc providing soft photons for Compton cooling.
Mathematically, the fully-advective branch would appear for $\Fsoft=0$,
which would require unphysical condition $f=1/[1-\eta (1-a)] > 1$.
Thus the necessary presence of the soft photons from the disc suppresses 
in our model the fully advective branch, present in other optically thin 
solutions, e.g. ADAFs (see Section~\ref{sec:adaf}).

The existence of the advection dominated branch of solutions was 
not found by WCZ, and the reason for that is given in 
Section~\ref{sect:whynotadv}.

The viscosity parameter $\alpha$ has a strong influence on the existence
and properties of the corona. For $\alpha$ lower than our assumed 0.3
the solutions are more limited in $\dot m$ and radius, the more so the
higher the advection coefficient $\delta$. For example, for $\alpha=0.03$
and $\delta=0.2$ only very spatially limited solutions exist.
For the same $\alpha$ but $\delta=0.5$ no solutions exit.

\subsection{Corona properties}

\subsubsection{Advection}
\label{sect:advec}

\begin{figure*}
\epsfxsize = \textwidth
\epsfbox[40 150 560 700]{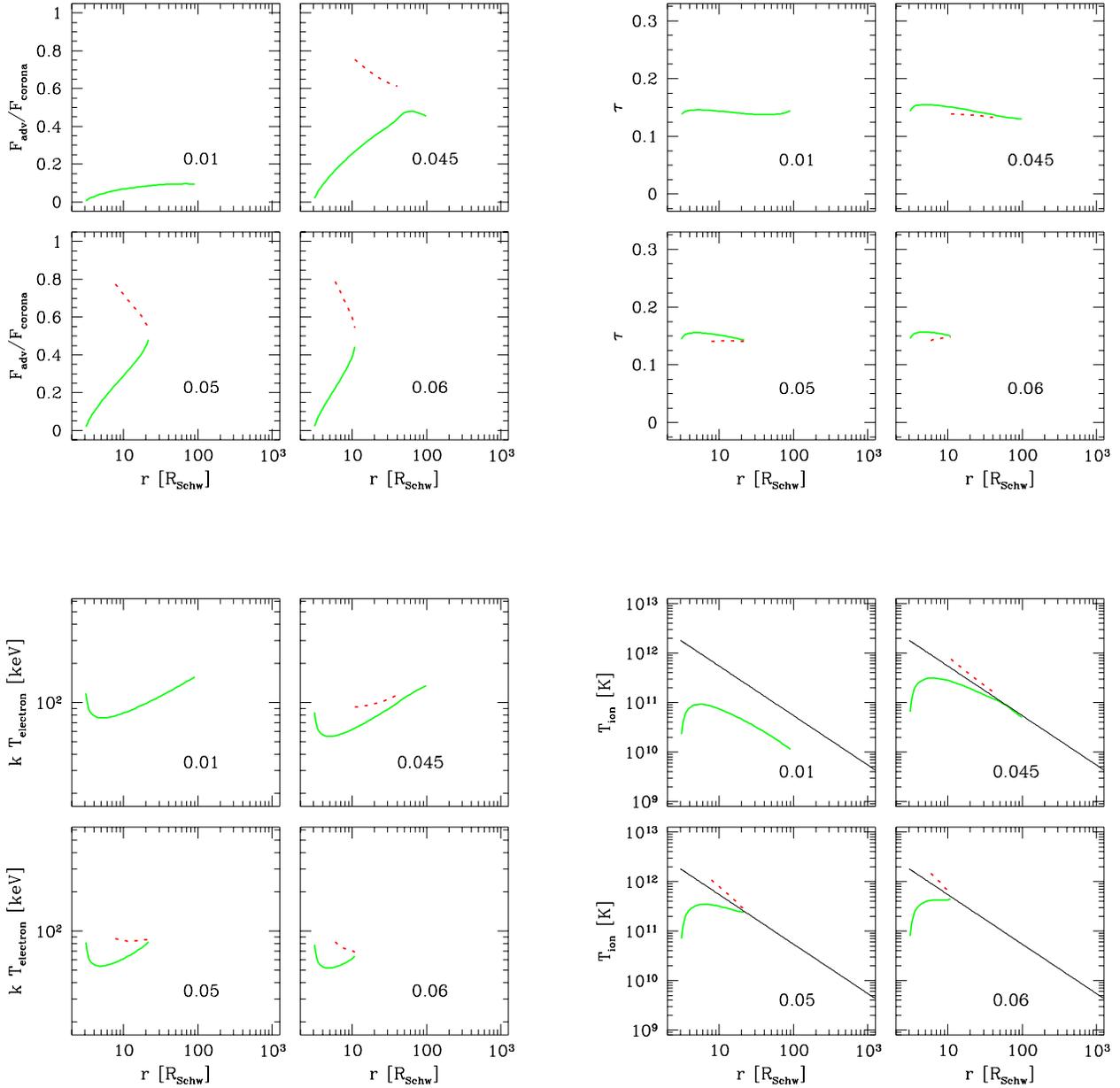}
\caption{The radial dependencies of the ratio of the advected energy flux to 
the total flux generated 
in the corona  (upper left panel),  optical depth (upper right panel),  
electron temperature (lower left panel), and ion temperature (lower right 
panel), for viscosity parameter $\alpha=0.3$, mass of the black hole 
$10 M_{\odot}$, and $\delta = 0.75$. The 
thin straight line in the lower right panel shows the virial temperature. 
The dashed line
shows advection--dominated solution whilst the solid line shows the solution
cooled predominantly by Compton cooling.
Labels inside panels refer to accretion rate.
\label{fig:adv}}
\end{figure*}

In Figure ~\ref{fig:adv} we show the ratio of the energy flux 
advected inwards with the coronal flow to the total dissipated flux
for $\delta=0.75$.
For low accretion rates only the radiatively cooled solution exists
and advection is not important. However, when $\dot m$ approaches
$\dot m = 0.1$, up to 40 per cent of the flux on this branch is carried 
by advection. This fraction depends on the radius. 

For high accretion rates ($\dot m \ga 0.04$) the second solution
appears. This solution is cooled mostly by advection and it is similar
to ADAF solutions.
Since in that case most of the energy (more than 90 per cent) is liberated in 
the corona the
accretion proceeds mostly through the corona itself.

\subsubsection{Ion temperature and the geometry of the corona}
\label{sect:whynotadv}

\begin{figure}
\epsfxsize = 80 mm 
\epsfbox[50 180 560 660]{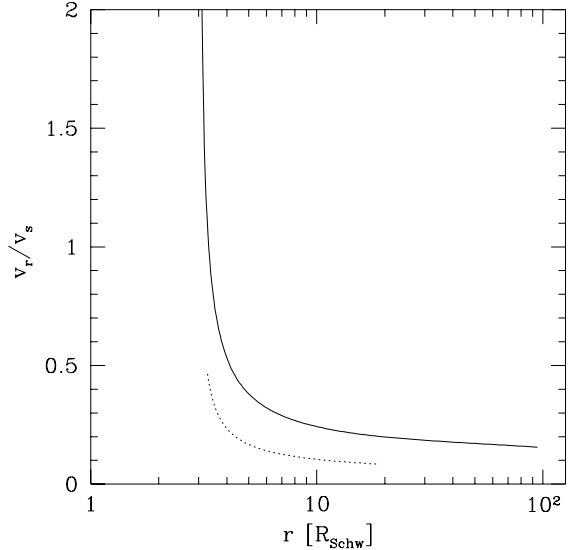}
\caption{The ratio of the radial velocity to the sound velocity in the corona
as a function of radius. The approximate constancy of $v_{\rm r}/v_{\rm s}$
for $r \ge 10\,\RSch$ means that the geometry of the accretion resembles
spherical accretion.
The viscosity parameter $\alpha = 0.3 $,  mass of the black hole 
$10 M_{\odot}$ and two values of
the accretion rate $\dot m$: $\dot m = 0.01$ (continuous line) and 
$\dot m = 0.001$ (dashed line).
\label{fig:vel}}
\end{figure}

The ion temperature decreases almost inversely with radius 
(see Figure ~5 in WCZ and formulae in Appendix C in that paper). 
The pressure scale height of the corona, defined as
\begin{equation}
H_{\rm P}=\left({ k \Ti R^3 \over G M m_{\rm H}}\right)^{1/2}
\end{equation}
increases almost linearly with radius 
and the ratio $H_{\rm P}/R$ 
is almost constant. The coronal accretion flow resembles actually a 
spherical accretion, similarly to the case of pure ADAF flows.
In such flows the ion temperature is of the order of the virial temperature and geometrical thickness of the flow is of the order of r. Sound velocity is close to the Keplerian velocity and is proportional to the radial velocity , where the proportionality coefficient is given by viscosity parameter $\alpha$. 
However, our coronal solutions are generally cooler and $H_{P}/R$ ratio, although constant, is not equal 1. Nevertheless, the dependence of the radial velocity on radius is quite similar to the case of ADAF. In Figure ~\ref{fig:vel} we plot the ratio of the radial to sound velocity, computed from the formula:  
\begin{equation}
v_{\rm r} = {f \dot M  \over 4 \pi R  \Sigmac}. 
\end{equation}
We see, that far from the marginally stable orbit $v_{r}/v_{s}$ ratio is of order of $\alpha$ and is almost constant throughout the disc. The flow is then moderately subsonic, depending on the value of the viscosity parameter. Close to the marginally stable orbit our coronal flow, as well as ADAF solution, become transonic and continue as a free fall onto a black hole.

In this paper we use a simplified description of the vertical hydrostatic
equilibrium and we have to check afterwards whether the solution can 
actually be in hydrostatic equilibrium. The approximate criterion 
is that the ion temperature should be smaller than the local virial
temperature. We see from Figure ~\ref{fig:adv} that $\Ti$ is usually 
lower than $\Tv$ on the radiative branch, but $\Ti > \Tv$ on the advective one.
The same problem refers to the corona 
height as a function of radius.
Since $H_{\rm P} /R = \sqrt{\Ti/\Tv}$, the ratio  $H_{\rm P}/R$ can be 
larger than 1, i.e.\ the corona can be (very) geometrically thick.
The reason for $\Ti$ exceeding $\Tv$ can be seen from 
Equation~\ref{equ:advfrac}:  $\Ti = (\Tv/\delta)\times(\Fadv/\Fc)$, i.e.\ 
$\Ti$ can approach and exceed $\Tv$ if advection is dominant.

The super-virial ion temperature is the reason why advection-dominated 
solutions were not found by WCZ. In that paper the vertical
structure was calculated much more carefully, assuming the hydrostatic 
equilibrium at the basis of the corona and allowing for the 
transonic vertical outflow from the corona.
These solutions automatically prohibited the
violation of the hydrostatic equilibrium at the basis of the corona. 
In that case the set of solutions for
increasing accretion rates simply
ended up as soon as the ion temperature reached the virial temperature 
(see Figure ~2 of WCZ and Section 3.1 and 4.1 therein).

The same problem was noted by Narayan \& Yi (1994) in the case
of pure ADAF solutions and expressed as the problem of the Bernoulli constant
being positive for ADAF.
Therefore, the model with a hot medium being in hydrostatic equilibrium 
does not offer correct solution beyond certain 
value of accretion rate.

\subsubsection{Electron temperature and the optical depth}

Since the density in the corona decreases outwards the efficiency of Coulomb 
interaction 
between the ions and electrons decreases as well. 
However, both the disc and the 
corona bolometric luminosities go down rapidly with the radius. 
Therefore the electron temperature, $\Te$, rises
outwards and the ratio of $\Ti/\Te$ decreases outwards.
The highest value of $\Te$ is of the order of $1.5 \times 10^9$ K, and it
depends only weakly on $\dot m$  and the viscosity parameter, $\alpha$. 
Such a universal value is an interesting
property of our model for lower accretion rates. At higher accretion rates, 
however, the outer radius of the corona contracts rapidly due to the advection
and the maximum corona temperature also rapidly drops down with an 
increase of the accretion rate.

The optical depth of the corona is practically independent of radius, and
very weakly dependent on other parameters; it is always between  
0.1 and 0.2 (see Appendix C in WCZ).

\section{Solutions with consistent $\bmath{\delta(\lowercase{r})}$}

\label{sec:diter}

\subsection{Corona structure}

\begin{figure}
\epsfxsize = 80 mm 
\epsfbox[20 300 570 720]{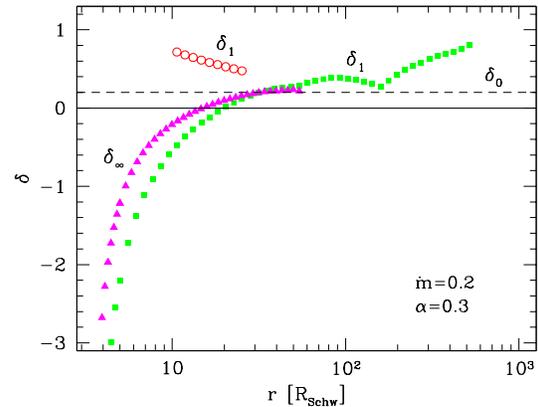}
\caption{Coefficient $\delta(r)$ computed from the solution for $\dot m=0.2$,
and initial $\delta_0=0.2$ (equation~\ref{equ:delta}). Solid squares
represent the radiative branch, while the circles show the advective
branch, after the first iteration ($\delta_1$). 
On the advective branch the computed $\delta$ is systematically larger
than the initial one which leads to the disappearance of this
branch, when solution with  consistent $\delta(r)$ is calculated.
Solid triangles ($\delta_{\infty}$) show $\delta(r)$ for the converged
solution, where only one branch remained.
\label{fig:delta}}
\end{figure}

The assumption that the advection parameter $\delta$ is constant as a function
of radius is not correct in general. As can be seen from 
equation~\ref{equ:delta},
$\delta$ is determined by radial derivatives of the ion temperature and
pressure, hence it can be a function of radius. 
Since the topology of solutions depends rather sensitively on
$\delta$ (Figure ~\ref{fig:xi}), we can expect significant changes of the 
topology if we consistently compute the $\delta(r)$ dependence.

\begin{figure*}
\epsfxsize = 0.8 \textwidth
\epsfbox[20 200 590 720]{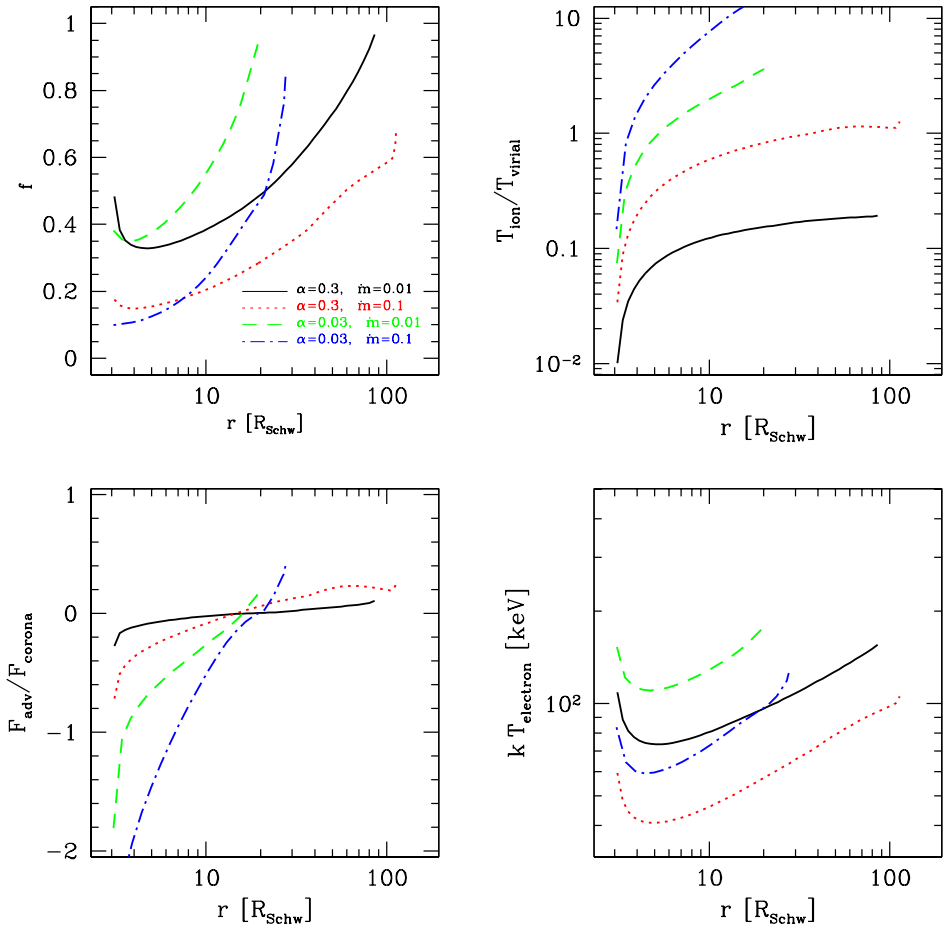}
\caption{Examples of the corona structure computed with consistent
solution for the advection coefficient $\delta(r)$. Two values of the
viscosity coefficient $\alpha$ are shown, with two values of $\dot m$ for 
each $\alpha$, as labelled. The solutions were computed for $10\ \MSun$, 
but the dependence on the BH mass is negligible.
\label{fig:iter}}
\end{figure*}

As a first step, we show in Figure ~\ref{fig:delta} the coefficient $\delta$ 
computed numerically from the solution obtained
for initial constant $\delta_0 = 0.2$ and $\dot m=0.2$ (labelled $\delta_1$)
For these parameters two
solution branches exist: with radiative and advective cooling dominant
(cf.\ Figure ~\ref{fig:xi}). On both branches the computed $\delta$ has
a strong radial dependence.  However, as Figure ~\ref{fig:xi} shows again, 
the radiative
branch is present for any $\delta$, therefore we do not expect significant
changes to its character, when solution with proper $\delta(r)$ is computed.
The same is not true for the advective branch: the higher the $\delta$, the
narrower the range of this branch's existence in the $r$--$\dot m$ plane. 
Since the computed $\delta$ is now significantly larger than the initial one, 
we can expect a decrease of the importance of the advective branch.

Proper self-consistent solutions describing the flow can in principle be
obtained either by solving radial differential equations containing explicitly
the advection term expressed as derivatives given by equation~\ref{equ:delta},
or by an iterative procedure correcting the distribution of $\delta(r)$ 
at each iteration step. First method was successfully applied 
by Chen, Abramowicz \& Lasota (1997) and Narayan, Kato \& Honma (1997)
to obtain global solutions for ADAF flow. We used the second method,
iterating solving the corona structure equations for a given $\delta(r)$
and computing the next approximation to $\delta(r)$.
Our algorithm is very similar to the method used
by Chen (1995): in order to find solution at a given radius $r$, we solve
the equations at $r$ and two auxiliary radii, $r-\Delta r$ and $r+\Delta r$
(assuming $\delta(r \pm \Delta r) = \delta(r)$, but the solution is 
insensitive to this particular condition). We then compute
the corrected $\delta$ from equation~\ref{equ:delta} and compute the
corrected structure. When convergence is achieved, we proceed to the next
radius.

Iterating  the solution for the advection-dominated branch quickly leads
to its disappearance.
The radiative branch shrinks somewhat ($\rmax$ decreases), especially
for $\dot m$ such that the advected fraction close to $\rmax$ was
substantial in the non-iterated solution. In other words, only the solutions
with rather low advective cooling survive.

Figure ~\ref{fig:delta} (points labelled $\delta_{\infty}$) shows an example 
of the iterated $\delta(r)$ dependence.
The iterated $\delta(r)$ is positive for larger radii, where advection is
a cooling process, but it changes sign  for smaller radii, as advection
becomes a locally heating process.
The same trend was obtained for optically thick discs calculated taking 
into account advection, departure from the Keplerian rotation and 
the transonic character of the flow close to 
the marginally stable orbit (Muchotrzeb \& Paczy\'{n}ski 1982; 
Abramowicz et al.\ 1988). The heating role of advection 
increases dramatically close to the marginally stable orbit since the energy 
dissipation there approaches zero.

In Figure ~\ref{fig:iter} we show examples of the converged solutions of the
corona structure as functions of $\dot m$ and $\alpha$. 
The corona is strongest at is outer edge, although $f$ is not always 1
at $r=\rmax$. Similarly
to solutions with constant $\delta$, the electron temperature is $\sim 100$
keV while the optical thickness is $\sim 0.15$. Advection is never important
as a cooling process. At small radii, $r \le 20\ \RSch$, $\delta$ changes
sign, so the advective flux contributes to  heating. In this region
the solution exist for $\dot m \ge 4\times 10^{-4}$, and up to $\dot m_{\rm c}
\sim 1$, i.e.\ the Eddington luminosity in the corona.
The topology of solutions represented on the $\dot m$--$\Sigma$ diagram
change as well (Figure ~\ref{fig:ADAF}): the solution forms a single branch
only.

The solutions are sensitive to the value of the viscosity parameter 
$\alpha$. For low $\alpha$ the solutions are generally rather limited
spatially. Moreover, for $\alpha \le 0.1$ the ion temperature 
strongly exceeds the 
virial temperature, so the solutions can hardly be considered physically
acceptable. 

The value of the mass of the central black hole influences 
the results only very slightly. Solutions for $M=10^{8}\ \MSun$ practically
overlap those for $M=10\ \MSun$. This is to be expected (see e.g.\ 
Bj\"{o}rnsson \& Svensson 1992 and references therein), since the only
dependence on the central mass in our model is through the temperature of 
the soft flux, which can affect the Compton amplification factor, but 
the dependence $A(T_0)$ is very weak for steep (soft) spectra.

\subsection{Radiation spectra}

\begin{figure*}
\epsfxsize = 0.9\textwidth
\epsfbox[20 400 600 710]{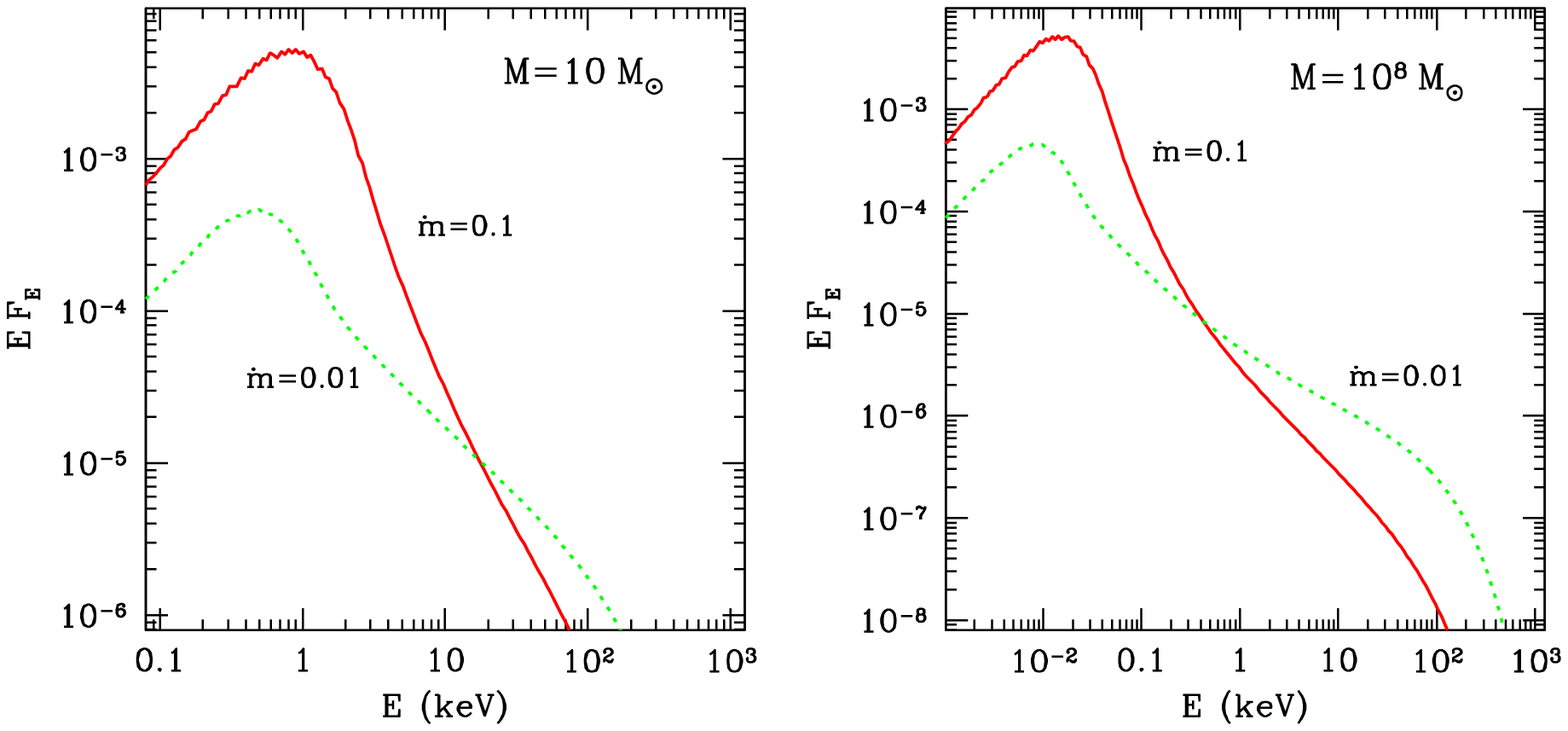}
\caption{The radiation spectra for $M=10\ \MSun$ (left panel) and
$M=10^8\ \MSun$ (right panel). Two values of the accretion rate are plotted,
as labelled, and the viscosity parameter $\alpha=0.3$. Corresponding
solutions of the corona structure are plotted in Figure ~\ref{fig:iter}.
\label{fig:spec}}
\end{figure*}

Spectra predicted by the model are rather soft and dominated by the disc
emission. Examples are plotted in Figure ~\ref{fig:spec} for parameters
characteristic for AGN and GBH. Generally, the power law components are harder 
for lower $\dot m$.

For accretion rates of order of $0.05\ \MEdd$, typically expected in Seyfert 
galaxies,
the amount of hard X-ray emission predicted by the model is negligible 
and the high energy index far
too steep,  so the present model does not offer a promising description. 
On the other hand, the original model of WCZ (with neglected advection) 
reproduced well the typical properties of AGN
(see Czerny et al.\ 1997 for radio quiet quasars and Seyfert galaxies, and
Kuraszkiewicz et al.\ 1999 for NLS1 galaxies; see also the 
corona parameters determined for MCG+8-11-11 by Grandi et al.\ 1998).

For the Galactic black holes the general trends are similar to those for 
AGN. For very low 
accretion rates ($\dot m \la 0.01$) the model predicts considerable
hard X-ray emission from the corona, although the spectra are somewhat
steeper than those resulting from the original model by WCZ  
(Janiuk \& Czerny 1999). 
Very steep spectra are predicted for $\dot m \ge 0.1$, i.e.\ corresponding
to the high or very high state of stellar black hole systems, while 
$\Gamma\sim 2$ in the observed spectra of e.g.\ soft X-ray transients
(\.Zycki, Done \& Smith 1998).

\section{Discussion}

In this paper we considered the accreting corona model.
Main features of the model and differences from other hot accretion flow
solutions are, firstly, the presence of a cool, optically thick disc 
supplying soft photons for Comptonization in the corona, and secondly, 
the condition for
(vertical) stratification of the flow into the two phases due to thermal 
instability.
Since in this model the accreting coronal plasma is hot, optically thin and 
geometrically thick,
it is necessary to include the radial energy transport (advection),
similarly to the optically thin ADAF solutions and optically thick,
slim discs.

The structure of solutions crucially depends on whether the advective flux
is solved for consistently -- i.e.\ the coefficient $\delta$ 
(Eq.~\ref{equ:delta}) is a function of radius --  or $\delta(r)$ is assumed 
constant, although there are certain properties of the solutions
independent of the $\delta$-prescription.

For a constant $\delta$, at a given radius (smaller than a certain maximum 
value) 
there is either one solution -- radiatively cooled through Compton process --
or two solutions, one radiatively cooled and the other advection-dominated.
For radii larger than the maximum one no coronal solutions exist.
Whenever two solutions are possible, there is always a critical value 
of the accretion rate for which the two solutions merge and no
solutions are found for higher accretion rates. 

Similar effect is seen when studying the radial structure of coronal solutions.
At very low accretion rates (below $\dot m \sim 4 \times 
10^{-4}$) there are no coronal 
solutions. At larger $\dot m$ the radiatively cooled solution
appears, which covers an inner part of the disc. This solution was
previously found by WCZ. The fraction
of the disc covered by the corona increases with the accretion rate but
the fraction of radiation emitted by the corona decreases. 
At even higher $\dot m$ (above $\dot m \approx 0.04$, depending on $\delta$) 
a second, 
advection--dominated solution emerges, if the corona structure equations
allow for the ion temperature to be higher than the virial
temperature.
%, as it is in the present paper.   
The two solutions merge at the outer edge of the corona. With further increase
of the accretion rate,
the region covered by the corona contracts rapidly, with no corona
present for $\dot m \ga 0.1$ (for the viscosity parameter 
$\alpha=0.3$). Smaller $\alpha$ leads to solution merger for
even lower accretion rates, as found previously by e.g.\ 
Abramowicz et al.\ (1995) and Bj\" ornsson et al. (1996).

The existence of the advection-dominated solutions was automatically 
excluded in the original formulation of the model by WCZ -- instead, 
they simply observed the disappearance of the single, radiation cooled
solution with an increase of the accretion rate due to the ion
temperature reaching the virial value.

When the $\delta(r)$ function is solved for, we obtain only one solution
branch. It is cooled by radiation i.e.\ with advective cooling negligible. 
In fact, advection changes sign for $r \le 20\ \RSch$ i.e.\ it acts as a 
heating process. The properties of this branch are similar to those of 
the radiatively cooled branch
obtained for a constant $\delta$: the solution exist for 
$\dot m > 4\times 10^{-4}$, and the corona is strongest at its outer edge.
For radii such that advection is a heating process there, this solution
can exist up to rather high $\dot m$, even formally exceeding $\MEdd$ 
in the corona. It disappears only when the temperature of the disc flux
reaches the (decreasing) electron temperature in the corona, but it is
strongly super-virial already for lower $\dot m$.

\subsection{Comparison with other hot, optically thin solutions}
\label{sec:adaf}

Our solutions show general similarity to other hot, optically thin disc 
solutions (SLE; Abramowicz et al.\ 1995; Chen et al.\ 1995; 
Bj\"{o}rnsson et al.\ 1996; Narayan \& Yi 1995; Zdziarski 1998;
see Kato et al.\ 1998 for general discussion).
There are also, however, certain 
important differences, as a direct consequence of the assumptions 
specific to our model.

When plotted on the $\dot m$--$\Sigma$ diagram (Figure ~\ref{fig:ADAF}),
our solutions for constant $\delta$ form two branches, merging for
certain maximum $\dot m$. Both the existence of the two branches and
their merging was found previously. However, as opposed to 
'conventional' ADAF solutions, strongly advection-dominated branch does
not appear in our model, due to substantial Compton cooling. 
For the same
reason our solutions disappear for lower viscosity, $\alpha \le 0.03$, while
no such effect is observed in the above mentioned solutions, 
where the adopted cooling mechanism can usually be made sufficiently 
inefficient (see below).

We obtain an  increase of the maximum allowed coronal $\dot m$, 
when we solve for the advective flux (i.e.\ $\delta(r)$). 
This has also been found previously by Chen (1995). 
Again, however, the advection-dominated branch
does not appear in our solution, and the solution disappears for low 
viscosity. None of these effects has appeared in the Chen (1995) work.

Hot discs cooled by comptonization of external soft photons were considered
by Zdziarski (1998), as a generalization of the SLE solution. However, the
source of soft photons was not specified in that work i.e.\ the soft flux
could implicitly be assumed to be small, if required. Therefore the general
properties of the solutions found were in close agreement with the 
previous solutions where bremsstrahlung was usually  assumed as the radiative 
cooling process. For example, the strongly advective branch is present in 
Zdziarski (1998) solution, in spite of the efficient radiative cooling.
The Compton parameter $y\equiv 4kT_{\rm e}/m_{\rm e} c^2 \,\tau$ is $\sim 1$ 
in that work (it is $\sim 0.1$ in our solutions), which indeed requires
very low $\Fsoft$ in order not to suppress the advectively-cooled branch.

Where an ADAF-like disc is assumed to co-exist with a cold, optically
thick disc (e.g.\ Esin et al.\ 1997), the transition radius is an adjustable
parameter. In our work the additional equation (Eq.~\ref{equ:xi}) resulting 
from the thermal instability condition closes the structure equations, thus
enabling us to compute the transition radius.

\subsection{Super-virial ion  temperature}

The ion temperature in an optically thin ADAF flow tends to be larger than
the local virial temperature and the Bernoulli constant for such a flow
is positive. Therefore such solutions are not possible without some kind 
of outflow (e.g. Narayan et al. 1997, Blandford \& Begelman 1999). 

The same problem affects the advection dominated branch of our coronal
solutions and the iterated solutions obtained for low $\alpha$.
Such a corona violates the assumption of the  
hydrostatic equilibrium. Coronal gas with ion temperature
exceeding the local virial temperature cannot flow in, as assumed in
our model or ADAF solutions. If strong outflow indeed developed,
no accretion would take place, switching off the energy source. 
A moderate transonic outflow from the
corona surface perpendicular to the disc does not provide a solution.
It was already included in the original formulation of the model
by WCZ and it did not prevent the disappearance of coronal solutions   
for accretion rates higher than $\dot m \sim 0.1$. A magnetic
wind  can provide a solution, provided that the outflow launched at certain
radius carries away {\it more\/} angular momentum than  what is locally
necessary for accretion to proceed
(Blandford \& Begelman 1999). However, at this stage such solutions 
are rather arbitrary.

\subsection{Corona formation at its outer edge}
\label{sec:form}

Our model does not provide a mechanism for disc evaporation. Instead, it
allows to check whether the corona, if formed, can exist in hydrostatic and
thermal equilibrium. The outer edge of our corona is therefore the maximum 
radius at which those conditions are satisfied. Since we did not consider 
two-dimensional flow, our transition from a bare disc to a coronal solution
is sharp. The transition would become smooth if the radial conduction were
included (e.g.\ Honma 1996), but this would require two-dimensional 
computations, since
the radial thickness of the transition zone is expected to be same 
as the vertical 
thickness (see discussion by Dullemond 1999). The coexistence of the bare disc
and the disc/corona at $\rmax$ does not contradict the analysis of
Dullemond \& Turolla (1999) since the coronal part is not strongly 
advection-dominated, with approximately half of the energy (or less) 
carried by advection and the remaining energy radiated away locally, as in SLE.

The mechanism leading to corona formation is still unspecified. It may be 
related either to disc instabilities, or magnetic phenomena. The coronal
solutions are generally within the standard disc radiation pressure 
instability zone (e.g.\ Janiuk \& Czerny 1999), but this does not seem 
a strong  argument in favour of the first possibility.

The transition radius, $\rmax$, in our models is generally
rather smaller than the outer radius of the ADAF flow in typical
applications, which is assumed $\sim 10^4\,\RSch$ (e.g. Esin et al. 1997).
Since our description 
of the flow applies in principle to the outer part of an ADAF solution, where 
the hot flow and the cold disc flow are assumed to overlap
(Esin et al.\ 1997),  it may mean that large ADAF radii would also be 
difficult 
to achieve if the disc/corona coupling was correctly
included. However, our result should be rather treated as an indication of a 
possible problem and not a definite answer. The model depends rather 
sensitively on the adopted description of the disc/corona coupling, expressed 
through Equation ~\ref{equ:xi2}, as was shown for a non-advective corona by 
Janiuk \& Czerny 2000. The simple change from the bremsstrahlung contribution
 from 2/3 (our assumption) to 3/7 (Krolik 1998) would only change the results 
quantitatively by radially expanding the region of coronal solutions but 
lowering the optical depth of the corona. However, better description of the 
transition region with disc evaporation included may change the results more 
significantly. It may also provide an explanation of the formation of ADAF 
part. Unfortunately, although some preliminary, partial results are available 
(e.g. Meyer \& Meyer-Hofmeister 1994, Dullemond 1999, 
R\' o\. za\' nska \& Czerny 1999), they are still not directly applicable to 
the models and further research is needed. It may also be
 true that the radial extension of the overlapping region will appear to be 
small, as suggested by Dullemond (1999), thus further complicating the 
prediction of location of the disc/ADAF transition.

\section{Conclusions} 
\begin{itemize}
\item Advection-dominated branch of coronal solutions does not represent 
a physically acceptable description of the flow.

\item Accreting corona solutions are predominantly Compton-cooled and,
for small radii, exist 
for all accretion rates larger than $\approx 4 \times 10^{-4}\MEdd$. 

\item  Spectral slopes predicted by accreting corona models for AGN and GBH 
are too steep in comparison with observations so the disruption of the 
innermost part of the disc or magnetically driven outflow seem to be required.
\end{itemize}

\bigskip

\section*{Acknowledgements}

This work was supported in part by grant 2P03D01816 of the Polish State 
Committee for Scientific Research.

\ \\
This paper has been processed by the authors using the Blackwell
Scientific Publications \LaTeX\  style file.

\end{document}